\documentclass[%
 aip,
 rsi,
 amsmath,amssymb,
 reprint,%
]{revtex4-1}

\usepackage{graphicx}
\usepackage{dcolumn}
\usepackage{bm}

\usepackage[utf8]{inputenc}
\usepackage[T1]{fontenc}
\usepackage{mathptmx}
\usepackage{etoolbox}
\usepackage{newtxtext}
\usepackage[varvw]{newtxmath}

\makeatletter
\def\@email#1#2{%
 \endgroup
 \patchcmd{\titleblock@produce}
  {\frontmatter@RRAPformat}
  {\frontmatter@RRAPformat{\produce@RRAP{*#1\href{mailto:#2}{#2}}}\frontmatter@RRAPformat}
  {}{}
}%
\makeatother
\begin{document}

\preprint{AIP/123-QED}

\title[A Bayesian approach to time-domain PDV analysis]{A Bayesian approach to time-domain Photonic Doppler Velocimetry analysis}
\author{J.~R.~Allison}
 \email{james.allison@firstlightfusion.com.}
\author{R.~Bordas}%
\author{J.~Read}
\author{G.~Burdiak}
\author{V. Beltr\'{a}n}
\author{N.~Joiner}
\author{H.~Doyle}
\author{N.~Hawker}
\author{J.~Skidmore}
\affiliation{ 
First Light Fusion Ltd., Unit 10 Oxford Pioneer Park, Yarnton, OX5 1QU, UK
}%

\author{T.~Ao}
\author{A.~Porwitzky}
\affiliation{%
Sandia National Laboratories, Albuquerque, New Mexico 87185, USA
}%
\author{D.~Dolan}
\affiliation{%
Physics Department and Institute for Shock Physics, College of Arts and Sciences, Washington State University, WA 99163, USA
}%
\author{B.~Farfan}
\author{C.~Johnson}
\author{A.~Hansen}
\affiliation{%
Sandia National Laboratories, Albuquerque, New Mexico 87185, USA
}%

\date{\today}

\begin{abstract}
Photonic Doppler Velocimetry (PDV) is an established technique for measuring the velocities of fast-moving surfaces in high-energy-density experiments. In the standard approach to PDV analysis, a short-time Fourier transform (STFT) is used to generate a spectrogram from which the velocity history of the target is inferred. The user chooses the form, duration and separation of the window function. Here we present a  Bayesian approach to infer the velocity directly from the PDV oscilloscope trace, without using the spectrogram for analysis. This is clearly a difficult inference problem due to the highly-periodic nature of the data, but we find that with carefully chosen prior distributions for the model parameters we can accurately recover the injected velocity from synthetic data. We validate this method using PDV data collected at the STAR two-stage light gas gun at Sandia National Laboratories, recovering shock-front velocities in quartz that are consistent with those inferred using the STFT-based approach, and are interpolated across regions of low signal-to-noise data. Although this method does not rely on the same user choices as the STFT, we caution that it can be prone to misspecification if the chosen model is not sufficient to capture the velocity behavior. Analysis using posterior predictive checks can be used to establish if a better model is required, although more complex models come with additional computational cost, often taking more than several hours to converge when sampling the Bayesian posterior. We therefore recommend it be viewed as a complementary method to that of the STFT-based approach.  
\end{abstract}

 \maketitle


\section{Introduction}
The velocity of fast-moving surfaces (such as a shock front or a projectile) is an important diagnostic for experiments in the field of shock physics\cite{Duvall:1977}. Two velocimetry methods are widely used, the Velocity Interferometer System for Any Reflector (VISAR\cite{Barker:1972, Hemsing:1979}) and Photonic Doppler Velocimetry (PDV\cite{Strand:2004, Strand:2006}); both use laser interferometry, but differ in the details of how velocity is inferred from the interference pattern. In the case of VISAR, which acts as a delay-leg interferometer, the variation in light intensity is directly proportional to the velocity of the surface, while for PDV the interference fringes are proportional to displacement and so must be differentiated with respect to time (usually by Fourier analysis). This requires a PDV system to have sufficient recording bandwidth to measure high velocities, although the advent of digital acquisition systems in excess of 10\,GHz has somewhat negated this issue. VISAR has the advantage of velocimetry resilience in the presence of rough surfaces (hence, `any reflector') while PDV can exhibit signal drop-out due to rough surfaces or multi-mode degraded fiber delivery. Conversely, PDV can be used to measure multiple velocities simultaneously, which is not possible with VISAR. Ultimately, since PDV is operationally simpler it has been widely adopted by the research community as an adjunct or replacement velocimetry method to VISAR\cite{Dolan:2024}. 

A full description of the standard approach to PDV analysis was given in the 2020 review by Dolan\cite{Dolan:2020}, which we briefly summarize here. Near-infrared laser light (typically $\lambda \approx 1550$\,nm) is reflected off a moving target and mixed with reference light (either from the same source or from a second laser) to generate a time-varying intensity signal at the receiver, the read-out of which is captured by a digital oscilloscope. A short-time Fourier transform (STFT) is then applied to the voltage versus time data to create a time-frequency representation of the signal (a spectrogram), from which dominant spectral components can be more easily identified and a velocity history extracted. The inferred velocity history is dependent on the choices made by the user that  include the form, duration and separation of the STFT window function, and the method to identify and characterize the time-varying spectrogram signal (typically using either peak finding, curve fitting or centroid analysis). The implicit time-scale is set by the STFT window function, which necessarily leads to temporal broadening of the signal that can obscure behavior at smaller time intervals. This is particularly evident at points of rapid acceleration/deceleration or discontinuities. The precision of the velocity measurement is ultimately limited by the sampling rate, signal noise fraction, and analysis duration\cite{Dolan:2010}, although other sources of error, such as laser speckle interference\footnote{We note that a technique based on polarization signal diversity technique has recently been developed for laser Doppler velocimetry \cite{Wang:2022, Wang:2024} which could be used in PDV to significantly reduce dropouts due to laser speckle noise. However, the relative cost of PDV (including GHz-bandwidth oscilloscopes and multiple channels) and extreme experimental conditions present significant barriers to use.}, are expected to contribute\cite{Ambrose:2017}.

Bayesian inference is a probabilistic approach to data analysis that has gained traction in many research disciplines because it provides a robust and easily interpretable method for determining, with uncertainty, model parameters from experimental data. Crucially, the use of Bayes' theorem allows the practitioner to naturally incorporate information from both prior knowledge and experimental evidence, optimizing the precision of the inferred parameter. In the domain of laser Doppler velocimetry (LDV), a closely-related diagnostic technique to PDV, Bayesian methods have been used with success to infer the dynamic behavior of vibrating solid objects\cite{Goggans:1999, Buell:2000} and fluid flows\cite{Fischer:2010, Penttinen:2015, Roncen:2018}. These methods often assume a simple model for the velocity behavior of the object under study, such as, in the case of acoustic oscillations, a time-varying sinusoid of constant phase and amplitude. This limits the number of model parameters and so greatly reduces the computational cost of numerically sampling the joint probability distribution. By contrast, PDV analysis poses unique challenges to the Bayesian approach because of the extreme physical conditions under study, including continuously time-varying supersonic velocity, the behavior of which is often unknown beforehand. Furthermore, PDV data are often highly-periodic and contain several thousand fringes in a single measurement, rendering a highly multi-modal probability distribution that is challenging to accurately sample using standard numerical approaches.  

In this paper, we use recent advancements in numerical sampling techniques for Bayesian inference to explore the potential of a time-domain approach to PDV analysis as a complementary method to the standard STFT-based approach. 
Specifically, we directly model the time-dependent oscilloscope signal and thereby avoid using the spectrogram, and associated user choices, as an analysis tool. 
We use a general parameterized time-series model for the velocity history, from which we generate synthetic PDV oscilloscope traces that can be compared with the observed data. 
Used within a Bayesian framework, we can automatically recover velocity histories from the data with uncertainties that are determined naturally from an input noise model.
Such an approach is susceptible to model misspecification and requires that we either capture possible errors in our model or that these are dealt with in data pre-processing. 
To avoid greatly increasing the number of required parameters in our model, we filter and normalize the data to remove any DC or low-frequency behavior (which provide no information about the velocity) and, subsequently, we construct a noise covariance model that captures the correlated behavior due to the filtering.
We verify this approach by recovering an input synthetic velocity history and then validate it against standard PDV analysis of a shock front propagating through a quartz sample from data obtained with the STAR two-stage light gas gun at Sandia National Laboratories\cite{Chhabildas:1982}. 

We structure this paper as follows: in Section\,\ref{sec:bayesian_pdv_analysis} we describe in detail our approach to Bayesian inference for PDV analysis, including verification against synthetic data, in Section\,\ref{sec:experimental_validation} we present results from analysis of PDV data from the STAR two-stage light gas gun, and in Section\,\ref{sec:concluding_remarks} we summarize our conclusions and recommendations.

\section{Bayesian PDV analysis}\label{sec:bayesian_pdv_analysis}

\subsection{Bayesian inference}

We use Bayesian inference to recover the velocity histories of moving targets from PDV data. This approach is advantageous over other techniques because it allows us to obtain probability distributions for the model parameters and derivative variables, and to include information from prior knowledge and experiments. Furthermore, by forward-modeling the system we are able to reliably capture the effect of systematic errors and uncertainties in the experiment. The disadvantages of this approach are that the result is highly dependent on the model and so any misspecification of the data, either in the signal or noise, will be propagated through as an unknown error. Knowledge of the limitations of the model are therefore important for interpreting any result obtained using this approach.

In most modeling problems, we are interested in inferring the posterior probability of the parameters $\boldsymbol{\theta}$ that describe our model $M$, given our data $D$. Using Bayes' theorem, we can express this in terms of the product of the \emph{likelihood function}, $P(D|\boldsymbol{\theta},M)$, and the \emph{prior probability distribution}, $P(\boldsymbol{\theta}|M)$, 
\begin{equation}
    \label{equation:bayes_theorem}
    P(\boldsymbol{\theta}|D,M) 
    = 
    \frac{P(D|\boldsymbol{\theta},M)P(\boldsymbol{\theta}|M)}{P(D|M)}
    \,,
\end{equation}
where the denominator, $P(D|M)$, known as the marginal likelihood (or model evidence), is independent of $\boldsymbol{\theta}$ and so can be absorbed into the normalization. In general, the posterior is not analytically tractable and so numerical methods, such as Monte Carlo sampling, are used to approximate the probability distribution as a function of the model parameters. From Eq.\,\ref{equation:bayes_theorem}, we can see that choosing the correct likelihood function and priors are critical in accurately representing our belief in the model parameters. In the following sections we present our choices of these for PDV analysis.

\subsection{Likelihood function}

The likelihood function, $\mathcal{L}(\boldsymbol{\theta}) = P(D|\boldsymbol{\theta},M)$, encodes information about the model parameters gained from the experimental data, updating our belief from the prior to the posterior probabilities. In the case of PDV analysis, the dataset is a voltage-time trace acquired with a digital oscilloscope. The ground-truth PDV signal is contaminated by random Gaussian noise due to components in the signal chain, which is dominated by photon noise from the reference laser (see Sections IV C and VII F of Dolan 2020\cite{Dolan:2020}). We therefore use a multivariate Gaussian likelihood function, given by
\begin{equation}
	\mathcal{L}(\boldsymbol{\theta}) = \frac{1}{\sqrt{(2\pi)^N\,|\mathbf{C}|}}\exp\left[-\frac{1}{2}(\boldsymbol{d}-\boldsymbol{\mu}(\boldsymbol{\theta}))^\mathrm{T}\mathbf{C}^{-1}(\boldsymbol{d}-\boldsymbol{\mu}(\boldsymbol{\theta}))\right],
\label{equation:likelihood_functions}
\end{equation}
where $\mathbf{C}$ is the matrix that encodes the covariance due to the noise, $\boldsymbol{d}$ is the vector (of length $N$) of sampled voltage data as a function of time, and $\boldsymbol{\mu}$ are the expected values of these data given by the model. The components of $\mathbf{C}$ are $\rho_{i, j}\sigma_{i}\sigma_{j}$, where $\sigma_{i}$ and $\sigma_{j}$ are the standard deviations due to noise at the $i$'th and $j$'th positions in the data, and $\rho_{i, j}$ are Pearson correlation coefficients. We describe how these are estimated and modeled in Section\,\ref{subsubsec:pdv_noise}. 

\begin{figure*}
    \includegraphics[width=1.0\textwidth]{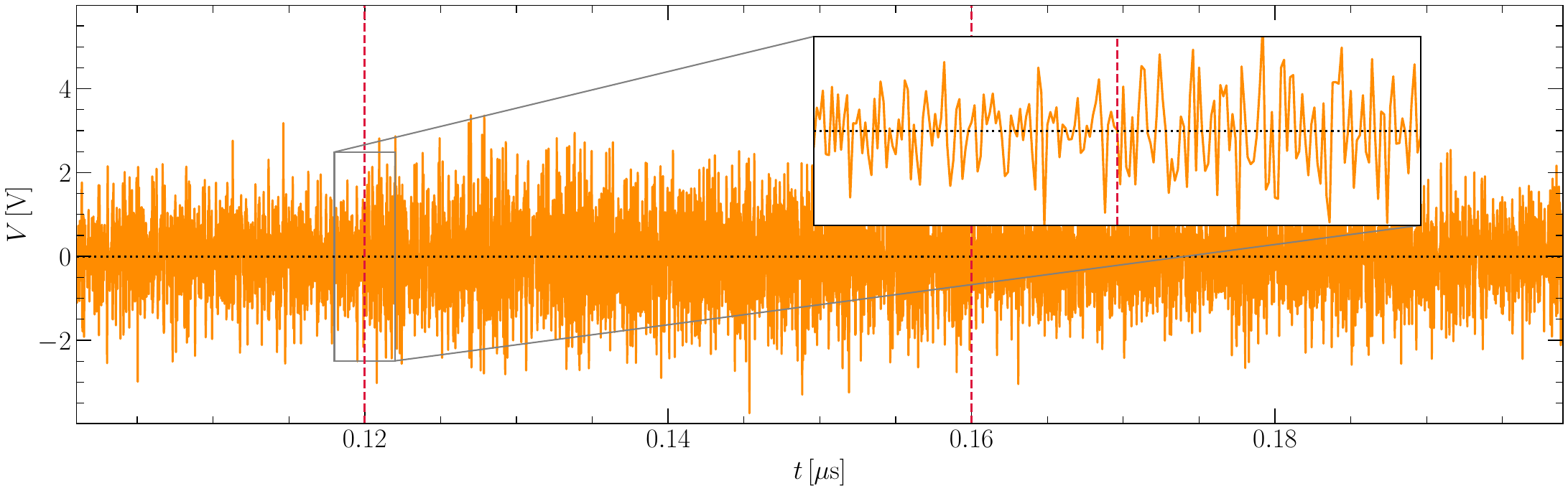}
    \caption{A synthetic PDV oscilloscope trace (voltage, $V$ versus time, $t$) generated using the models discussed in Section\,\ref{subsec:models} and used for verification of our Bayesian method (see Section\,\ref{subsec:verification_against_synthetic_data}). The vertical red lines correspond to the start and end times, at 0.12 and 0.16\,$\mu$s respectively, of the target signal. Although the ground-truth PDV signal is sinusoidal, the presence of noise renders visual identification of the fringes challenging (see inset); hence the need for techniques such as STFT analysis or Bayesian forward-modeling to accurately infer the velocity history from the data.}
    \label{figure:synthetic_verification_data_pdv_trace}
\end{figure*}

\subsection{Models}\label{subsec:models}

\subsubsection{Velocity history}\label{subsubsec:velocity_history}

We model the velocity history of a single component using a parameterized one-dimensional series based on piecewise linear components. Parameters consist of an initial velocity, $v_{0}$, and $n$ accelerations $\dot{v}_i$ over each subsequent time interval $t_{i} - t_{i-1}$. By parameterizing the series in terms of acceleration, we do not need to adjust the prior distributions with respect to temporal resolution. The velocity at time $t \in (t_{j-1},~t_{j})$ is then given by 
\begin{equation}
    v(t) = v_{j-1} + (v_{j} - v_{j-1})\,\frac{t - t_{j-1}}{t_{j} - t_{j-1}},
\end{equation}
where
\begin{equation}
v_{j} = v_0 + \sum_{i=1}^{j}{\dot{v}_{i}\,(t_{i} - t_{i-1}}).
\end{equation}
The number of acceleration parameters depends on the desired time resolution but are constrained by the computational complexity of sampling the parameter space for inference. The start and end times of the series could be included as inferred parameters, but we found that model misspecification of real experimental data leads to concentration of fitting in high signal-to-noise regions (see Section\,\ref{sec:experimental_validation}). Therefore, these are fixed to reasonable values that enclose the PDV signal based on visual inspection of the spectrogram.

This model has the advantage of simplicity and intepretability, but forces linear behavior over time scales smaller than $\Delta{t}_i$ and is not differentiable at component intersections. This behavior could be avoided by choosing alternative models that are constructed from linear combinations of a finite set of orthogonal basis functions, which include, for example, the Hermite Polynomials or Karhunen-Lo\'{e}ve (KL) expansion of a Gaussian Process (GP) covariance function\cite{LeMaitre:2010, Rasmussen:2006}. Likewise, systems that are known to include discontinuities in the velocity history could be explicitly modeled by incorporating a finite number of parameterized step functions that multiply the time series. Such models are beyond the scope of this work but should be considered in future implementations. 

\subsubsection{PDV signal}
\label{subsubsec:pdv_signal}

We generate a synthetic PDV signal by modeling the interference fringe pattern between light from the target and reference lasers. The time-dependent beat frequency of this interference signal is given by 
\begin{equation}
    f(t) \equiv \frac{\mathrm{d}\phi(t)}{\mathrm{d}t} = |\nu_\mathrm{T} - \nu_\mathrm{R} + 2\,\nu_\mathrm{T}\,\beta^\ast(t)|,
\end{equation}
where $\phi$ is the phase angle, $\nu_\mathrm{T}$ and $\nu_\mathrm{R}$ are, respectively, the target and reference laser frequencies, and $\beta^\ast(t)$ is the ratio of the apparent velocity, $v^\ast(t)$, to the speed of light in vacuum. Uncertainties in the PDV laser frequencies could be included as a component of the model, but are expected to be at the level of one part in a million\cite{Dolan:2020} and so we do not include this in our analysis. For a full discussion of the definition of apparent velocity please see section\,VI\,B of Dolan (2020)\cite{Dolan:2020}; here, for the simple reflective target surfaces considered in this work, the apparent velocity is related to the lab-frame velocity, $v(t)$, by
\begin{equation}
    v^\ast(t) = n_0\,v(t),
\end{equation}
where $n_{0}$ is the refractive index ahead of the target. 

For near infrared lasers ($\lambda \approx 1550$\,nm) that are typically used in PDV, changes in beat frequency are related to changes in the apparent velocity by
\begin{equation}
    \left[\frac{\Delta{f}}{1.3\,\mathrm{GHz}}\right] \approx G(v^\ast)\,\left[\frac{\Delta{v^\ast}}{1\,\mathrm{km\,s^{-1}}}\right],
\end{equation}
where $G(v^\ast)$ can be $\pm1$, given by
\begin{equation}
    G(v^\ast) = 2\,H(\nu_\mathrm{T} - \nu_\mathrm{R} + 2\,\nu_\mathrm{T}\,\beta^\ast) - 1.
\end{equation}
$H$ is the Heaviside step function, which has the property of being zero for arguments less than zero, and unity otherwise. Therefore, for fast-acquisition oscilloscope bandwidths $\gtrsim 10$\,GHz, changes in apparent velocities $\gtrsim 8$\,km\,s$^{-1}$ can be measured. If $\nu_\mathrm{T} > \nu_\mathrm{R}$, then $G(v^\ast) = 1$ for all positive (that is, approaching) apparent velocities and the diagnostic is known as up-shifted PDV. Alternatively, for down-shifted PDV, the target and reference laser wavelengths are chosen such that $\nu_\mathrm{T} < \nu_\mathrm{R}$ and so $G(v^\ast)$ switches sign within the range of positive apparent velocities. This doubles the range of measurable velocities within the available oscilloscope bandwidth, but with ambiguity about the frequency-to-velocity mapping (see figure 4 of Dolan 2020\cite{Dolan:2020}).

Once we have modeled the beat frequency as a function of time, the resulting PDV signal is given by
\begin{equation}
    V(t) = \hat{V}(t)\,\sin\left(\phi_0 + \int_{t_0}^t{f}(t^\prime)\,\mathrm{d}t^\prime\right),
\end{equation}
where $\hat{V}$ is the amplitude (dependent on the unknown time-varying intensities of the target and reference lasers), and $\phi_0$ is the unknown phase at $t = t_0$. We cumulatively evaluate the integral numerically using the composite trapezoid method. We could model the amplitude as a time series, using the models described above, but this would not provide us with further information about the velocity history of the target. Alternatively, we can attempt to capture and factor out the amplitude during a pre-processing step, thereby being left with only the frequency-dependent component of the signal. However, this assumes that only a single coherent component exists in the PDV signal. We discuss this further in Section\,\ref{sec:experimental_validation}. In the case of the unknown phase offset, we simply include $\phi_0$ in the model as a nuisance parameter. 

\renewcommand{\arraystretch}{1.2}
\begin{table}
    \centering
    \caption{Model parameter values used to generate a synthetic PDV dataset for verification of the method.}
    \label{table:verification_model_parameters}
    \begin{tabular}{cccc}
        \hline
        \hline
        Model & Parameter & Units & Value \\
        \hline
        \hline
        Velocity history & $v_{0}$ & km\,s$^{-1}$ & 25 \\
        \vdots & $\dot{v}_{1}$ & km\,s$^{-1}$ per $\mu$s & 0 \\
        & $\dot{v}_{2}$ & \vdots & 0 \\
        & $\dot{v}_{3}$ & & -20 \\
        & $\dot{v}_{4}$ & & -50 \\
        & $\dot{v}_{5}$ & & -100 \\
        & $\dot{v}_{6}$ & & -100 \\
        & $\dot{v}_{7}$ & & -100 \\
        & $\dot{v}_{8}$ & & -50 \\
        & $\dot{v}_{9}$ & & -20 \\
        & $\dot{v}_{10}$ & & 0\\
        & $\dot{v}_{11}$ & & 0 \\
        & $\dot{v}_{12}$ & & 20 \\
        & $\dot{v}_{13}$ & & 50 \\
        & $\dot{v}_{14}$ & & 100 \\
        & $\dot{v}_{15}$ & & 50 \\
        & $\dot{v}_{16}$ & & 20 \\
        & $\dot{v}_{17}$ & & 0 \\
        & $\dot{v}_{18}$ & & -10 \\
        & $\dot{v}_{19}$ & & -20 \\
        & $\dot{v}_{20}$ & & -50 \\
        \hline
        Phase offset & $\phi_{0}$ & deg & 0 \\ 
        \hline
        \hline
    \end{tabular}
\end{table}

\renewcommand{\arraystretch}{1.2}
\begin{table*}
    \centering
    \caption{Priors used in this work for each model parameter.}
    \label{table:prior_distributions}
    \begin{tabular}{ccccc}
        \hline
        \hline
        Model & Parameter & Units & Prior distribution & Prior hyperparameters \\
        \hline
        \hline
        Velocity history & $v_{0}$ & km\,s$^{-1}$ & Gaussian & $\mathrm{mean} = 25, \mathrm{std} = 1$ \\
        & $\dot{v}_{i}$ & km\,s$^{-1}$ per $\mu$s & Gaussian & $\mathrm{mean} = 0, \mathrm{std} = 100$ \\  
        \hline
        Phase offset& $\phi_{0}$ & deg & Uniform & $\mathrm{min} = -180, \mathrm{max} = 180$ \\ 
        \hline
        \hline
    \end{tabular}
\end{table*}

\subsubsection{PDV noise}\label{subsubsec:pdv_noise}

We model the noise in the PDV data as a multivariate Gaussian distribution with zero mean and covariance $C_{i, j} = \rho_{i, j}\,\sigma_{i}\,\sigma_{j}$, where $\sigma_{i}$ and $\sigma_{j}$ are the standard deviations at the $i$'th and $j$'th positions in the data, and $\rho_{i,j}$ is the Pearson correlation coefficient. We could include these parameters in the inferred model, but this would greatly increase the complexity of the inference problem. Instead we estimate the noise \emph{a\,priori} using the following summary statistics based on prior assumptions:

\paragraph*{Standard deviations:} Under the assumption of homoskedastic (that is, constant variance) noise, we can estimate a single value for the standard deviation across the oscilloscope trace. We identify a signal-free region in frequency space from a spectrogram of the data, then apply an appropriate noise-only bandpass filter to remove the PDV signal. In the case of the experimental validation data used in Section\,\ref{sec:experimental_validation}, we used beat frequencies between 2 and 5\,GHz as representative of signal-free data (see the spectrograms in Fig.\,\ref{figure:star_endor_pdv_spectrograms}). We then estimate the standard deviation due to noise using the median absolute deviation (MAD) across the filtered trace. In the presence of any residual signal or artifacts, MAD is a more reliable estimator of the variability due to noise than the standard deviation estimator. Finally, because the noise-only filter acts to reduce the true noise, we apply a correction factor to the standard deviation that is estimated using a Monte Carlo sample of 1000 filtered synthetic Gaussian noise traces.  

\paragraph*{Correlation coefficients:} With regard to the correlation coefficients, we assume that all correlated noise is the result of bandpass filtering due to the oscilloscope bandwidth and subsequent data processing. We identify the frequency band of the oscilloscope trace from the spectrogram and then estimate the correlation coefficients from a Monte Carlo sample of 1000 bandpass filtered synthetic Gaussian-noise traces. 

We demonstrate the validity of this model in Appendix\,\ref{app:demonstration_of_noise_model}. Time series are shown for Gaussian noise generated with and without a covariance matrix determined using the method above for a 1\,GHz low-pass filter. The resulting spectrograms demonstrate that the covariance matrix correctly models the noise correlation due to the filter.

\begin{figure*}
    \includegraphics[width=1.0\textwidth]{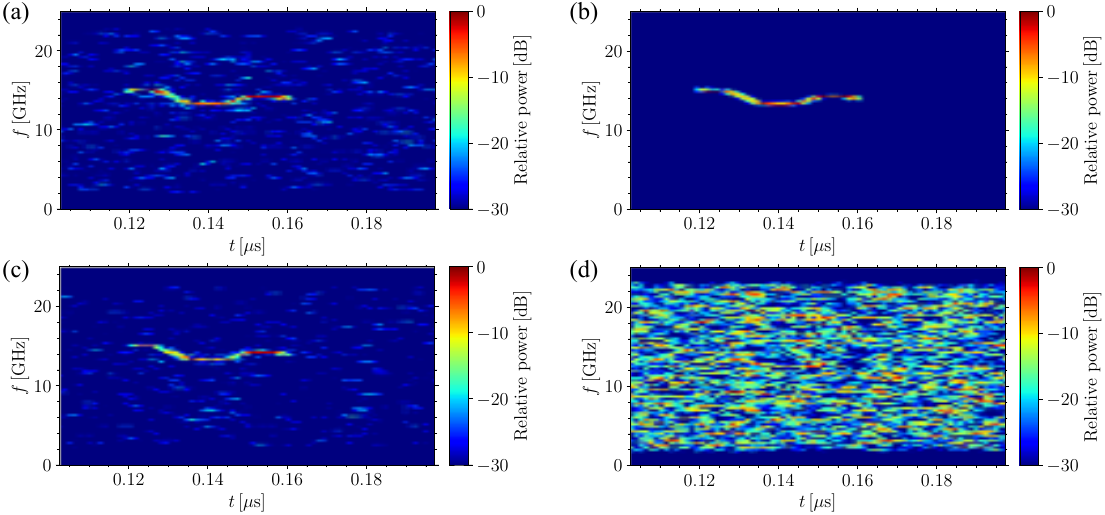}
    \caption{Spectrograms (beat frequency, $f$ versus time, $t$) generated from (a) the synthetic PDV data shown in Fig.\,\ref{figure:synthetic_verification_data_pdv_trace}, (b) the inferred posterior distribution of model PDV signals, (c) the inferred maximum \emph{a posteriori} (MAP; that is, most probable) model and (d) the corresponding residual (MAP model subtracted from the data). These visually demonstrate that the synthetic signal is recovered by the Bayesian forward-modeling method. In the case of the MAP spectrogram, we added random noise drawn from our PDV noise model for visual comparison with the data. The noiseless bands at $f < 2$ and $f > 23$\,GHz are the result of bandpass filtering during pre-processing.}
    \label{figure:synthetic_verification_spectrograms}
\end{figure*}

\begin{figure*}
    \includegraphics[width=1.0\textwidth]{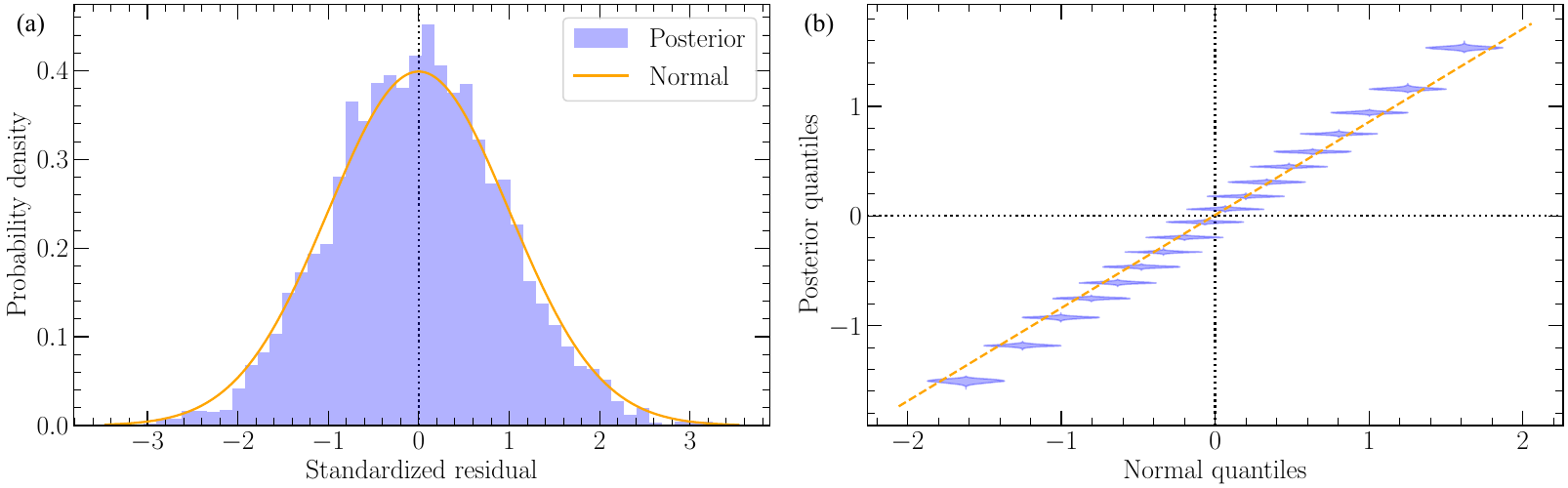} 
    \caption{(a) The posterior distribution of standardized residuals (PDV model subtracted from data) and (b) resulting quantile-quantile (QQ) plot with respect to the normal distribution. The synthetic noise is Gaussian and so residual deviation from a normal distribution is a measure of the error in recovering the synthetic PDV signal; this is quantified in the QQ plot by the degree of deviation from linear.}
    \label{figure:synthetic_verification_posterior_analysis}
\end{figure*}

\begin{figure*}
    \includegraphics[width=1.0\textwidth]{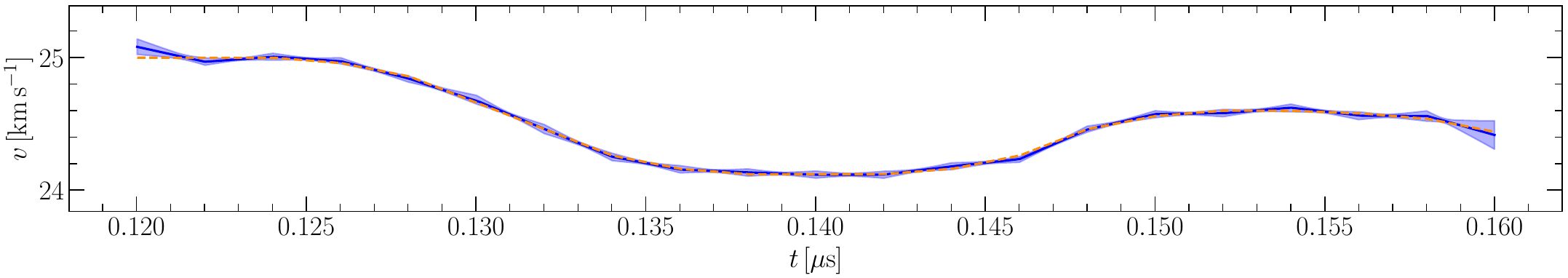}   
    \caption{The posterior velocity history (velocity, $v$ versus time, $t$) inferred from the synthetic PDV data shown in Fig.\,\ref{figure:synthetic_verification_data_pdv_trace}. The blue solid line and envelope, respectively, denote the median and 95\,\% credible interval over the sample distribution of velocity curves. The orange dashed line denotes the injected model from which the synthetic data were generated. The "bow tie" shape is a result of using a piecewise linear model for the time series.}
\label{figure:synthetic_verification_posterior_velocity_history}
\end{figure*}

\subsection{Implementation}
\label{subsec:implementation}

We implement all data ingestion, processing and modeling in the \texttt{Python} programming language. We use the \texttt{Bilby} Bayesian inference library\cite{Ashton:2019}, \footnote{Documentation for \texttt{Bilby} can be found at https://bilby-dev.github.io/bilby/} with the \texttt{Dynesty} Nested Sampling (NS) package\cite{Speagle:2020}, \footnote{Documentation for \texttt{Dynesty} can be found at https://dynesty.readthedocs.io} to recover an approximation of the posterior probability distribution for the model parameters. NS\cite{Skilling:2006} is an alternative approach to the commonly used Markov Chain Monte Carlo (MCMC) methods that step a chain of samples through the parameter space; instead, a one-dimensional integral is calculated over slices of the prior volume that are constrained by nested iso-likelihood contours. Although originally developed with the goal of estimating the model evidence, NS also produces the sampled posterior distribution that can be used for parameter estimation. Given the number of parameters and inherently multi-modal nature of the PDV likelihood, we use the random walk and multi-modal bounding options for live point replacement sampling (LRPS), with 1500 live points used for robust posterior sampling (for a review of these methods please refer to Buchner 2023\cite{Buchner:2023}). This produces a chain of sample points that can be used to estimate the joint posterior distribution of the model parameters, and hence recover the PDV velocity history. The run time is dependent on the number of live points and the complexity of the posterior distribution; PDV analysis in the voltage-time space is a difficult inference problem and so the NS algorithm can take many hours to converge with more than 1000 live points.  

\subsection{Verification against synthetic data}
\label{subsec:verification_against_synthetic_data}

\subsubsection{Synthetic data generation}
\label{subsubsec:synthetic_data_generation}

To verify that our method can accurately recover velocity histories from PDV data, we tested it on a synthetic dataset that is similar to the data used for experimental validation in Section\,\ref{sec:experimental_validation}. We generated a synthetic oscilloscope trace over an interval of $0.1$ -- $0.2$\,$\mu$s, with a sample interval of 0.02\,ns that is typical of multi-GHz oscilloscopes. We added a PDV signal with start and end times of 0.12 and 0.16\,$\mu$s, respectively, amplitude $\hat{V} = 1$\,V, phase offset $\phi_{0} = 0$\,deg, and laser wavelengths $\lambda_\mathrm{T} = 1549.219$\,nm and $\lambda_\mathrm{R} = 1548.945$\,nm. These wavelengths correspond to down-shifted PDV with a "bounce" at an apparent velocity of $v^{\ast} = 26.5$\,km\,s$^{-1}$. 

The velocity history was modeled using the piecewise linear model described in Section\,\ref{subsubsec:velocity_history}, with $v_{0} = 25$\,km\,s$^{-1}$ and twenty $\dot{v}_{i}$ parameters  (see Table\,\ref{table:verification_model_parameters}) that generate arbitrary, but typical, varying behavior of a propagating shock front in a solid diagnostic block. This was then converted to the apparent velocity using a refractive index of $n_{0} = 1.528$, equal to that of the quartz sampled used in the experiment described in Section\,\ref{sec:experimental_validation}. Finally, we added noise with unity standard deviation and applied a bandpass filter between 2 and 23\,GHz, to simulate the AC coupling and bandwidth limit of the scope, respectively. The resulting synthetic oscilloscope trace is shown in Fig.\,\ref{figure:synthetic_verification_data_pdv_trace}  and the spectrogram in the top left hand panel of Fig.\,\ref{figure:synthetic_verification_spectrograms}. 

\subsubsection{Model prior selection}
\label{subsubsec:model_prior_selection}

It is evident from visual inspection of the synthetic oscilloscope trace that this is a challenging inference problem; we are attempting to infer a model from noisy data that describes the time-varying behavior of a periodic signal with several thousand cycles. The likelihood function is inherently multi-modal because frequency aliases, corresponding to higher velocities, will produce equally good representations of the data. Likewise, the signal periodicity will generate several local likelihood maxima that are likely to prove challenging for numerical samplers. Given these challenges, we adopted Gaussian prior distributions for our velocity model parameters that are weakly informative and therefore able to regularize the inference against finding aliased modes with physically-undesirable velocities. In particular, the prior mean and standard deviation for the initial velocity, $v_{0}$, is chosen based on inspection of the signal in the spectrogram. This is further helped by including the phase offset parameter, which shifts the model to better match the true periodic PDV signal. We summarize the prior distributions for each model parameter in Table\,\ref{table:prior_distributions}.

\subsubsection{Verification results}
\label{subsubsec:verification_results}

In Fig.\,\ref{figure:synthetic_verification_spectrograms} we compare spectrograms for the synthetic data with those obtained from the inferred model velocity history. In the top-right panel we show the full posterior spectrogram, which is generated by stacking spectrograms for each set of model parameters sampled from the posterior probability distribution. Visually, this is comparable to the signal seen in the data. We also show the maximum \emph{a posteriori} (MAP) spectrogram, with noise added from our model, which again is visually consistent with the data. The corresponding MAP residual is shown and consistent with noise, again demonstrating that the synthetic signal is recovered by the method.

This is further demonstrated in Fig.\,\ref{figure:synthetic_verification_posterior_analysis}, where we compare the posterior distribution of residuals in the PDV trace (difference between model and data) with the normal distribution. The synthetic noise is Gaussian and so any deviation from the normal distribution is a metric for errors in recovering the synthetic signal. Quantitative analysis is done using the quantile-quantile (QQ) plot\cite{Wilk:1968}, which graphically assesses whether the posterior is plausibly consistent with a normal distribution. The linearity of the points suggest that the residuals are indeed drawn from a normal distribution.  

As a final assessment of the verification, in Fig.\,\ref{figure:synthetic_verification_posterior_velocity_history} we compare the inferred posterior velocity history with the injected model. They are consistent within the 95\,\% credible interval of the posterior, demonstrating that the injected model is recovered with high confidence.

\section{Experimental validation}
\label{sec:experimental_validation}

\begin{figure*}
    \centering
    \includegraphics[width=0.925\textwidth]{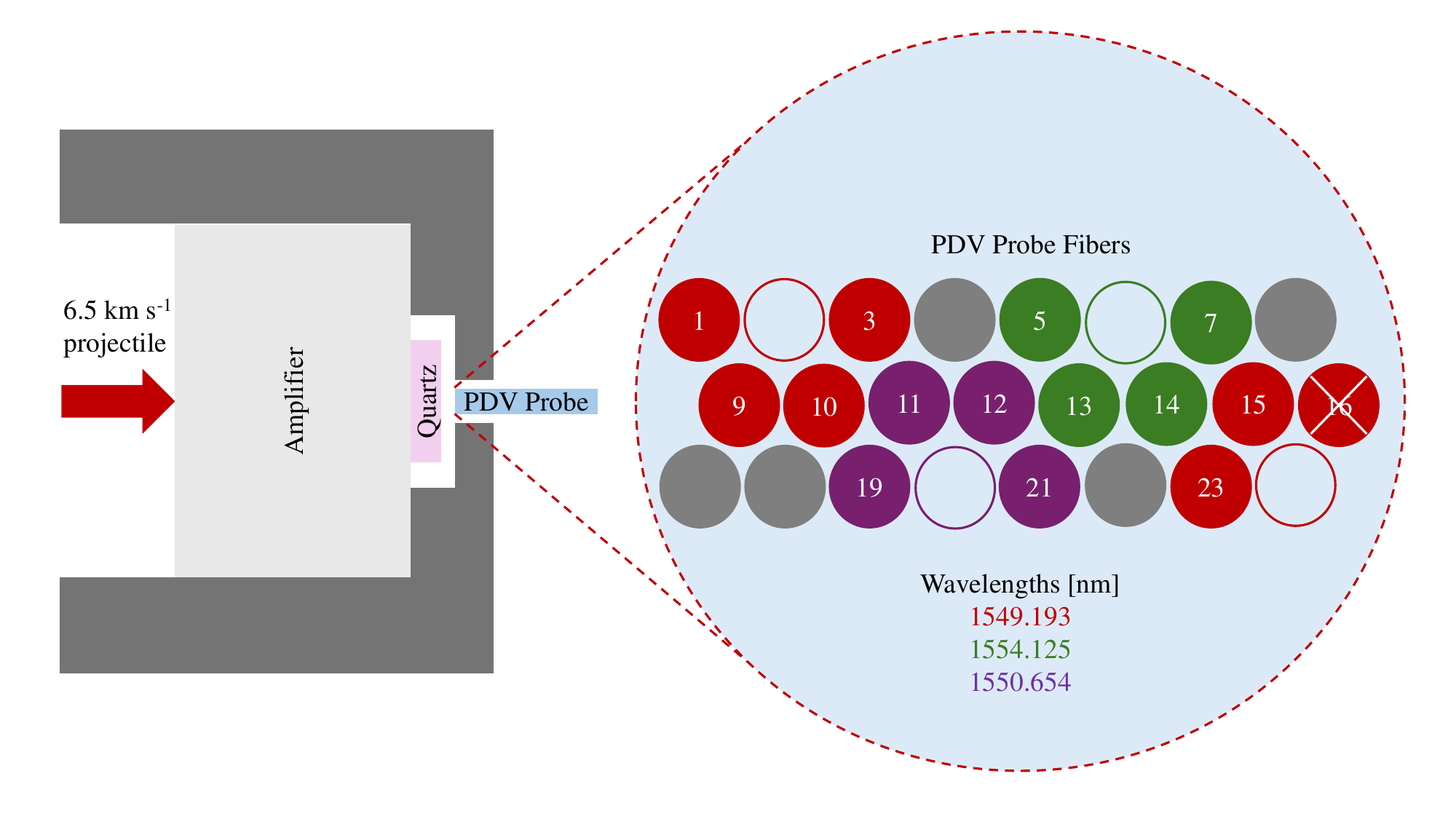}\\
    \caption{Schematic (not to scale) of the experimental set up used to provide a validation dataset for our Bayesian approach to PDV analysis. A 17\,mm diameter tantalum projectile was launched at a velocity of 6.5\,km\,s$^{-1}$ using the STAR two-stage light-gas gun at Sandia National Laboratories\cite{Chhabildas:1982}, which drove a pressure-amplified planar shock into a 3.5\,mm diameter quartz sample at velocities $\approx25$\,km\,s$^{-1}$. The velocity history of the shock wave was measured using a bespoke multi-point PDV probe containing an array of 24 fibers, each separated by 125\,$\mu$m, and arranged in the configuration shown in the right-hand insert. Four fibers (empty circles) were used to send the laser light to the quartz sample and 15 (numbered circles) to return the reflected light from the shock front. Fiber 16 (crossed out) was broken and so did not return any usable data. The colors denote fiber light of the same wavelength, which are given in Table\,\ref{table:experimental_parameters}.}
    \label{figure:star_endor_experimental_setup}
\end{figure*}

\subsection{Experimental setup}

Validation of our Bayesian approach to PDV analysis was achieved by comparing the inferred velocity histories from a real experimental dataset with those obtained using the standard STFT-based method. To this end, we used data from a recent experiment by Skidmore et al.\cite{Skidmore:2025} with the STAR two-stage light-gas gun (2SLGG) at Sandia National Laboratories\cite{Chhabildas:1982}. Skidmore et al.\ used down-shifted PDV to measure the velocity of a shock driven into a 3.5\,mm diameter quartz sample by a 17\,mm diameter tantalum projectile launched at 6.5\,km\,s$^{-1}$ using the 2SLGG. This was carried out as part of a commissioning campaign for a novel hydrodynamic pressure amplifier developed at First Light Fusion (FLF) as a platform for low-cost Equation-of-state (EoS) measurements at terapascal pressures. The geometry of the pressure amplifier (to be presented in a forthcoming paper by Skidmore et al.\cite{Skidmore:2025}) is such that the shock wave is split and recombined to form a Mack stem at the output. The resulting planar shock front is sufficient to uniformly compress the quartz sample, situated at the output, to pressures exceeding 1\,TPa across a 1\,mm diameter region (equivalent to velocities $\approx 25$\,km\,s$^{-1}$); similar schemes have been fielded previously on laser platforms\cite{Swift:2006}. 

To test the amplifier performance, a bespoke multi-point PDV probe of 24 optical fibers was placed in three rows extending 875\,$\mu$m across the quartz sample as shown in Fig.\,\ref{figure:star_endor_experimental_setup}. Four fibers were used to send laser light to the sample, and 15 to return the reflected signal (although one return fiber was broken and so did not provide usable data). The incident light used an NKT Koheras BASIK laser source at 40\,mW connected to a 4 way light splitter. Each splitter output was connected to an independent attenuator to regulate the incident power so as to obtain a -30\,dBm of return light in the PDV probe, measured using a power meter in the return fiber. Multiple target and reference wavelengths were used to minimize the cross-talk between adjacent channels; we summarize these in Table\,\ref{table:experimental_parameters}. 

Skidmore et al.\ carried out PDV analysis using the standard STFT approach, obtaining velocity histories that match the performance requirements of the amplifier. However, they noted that this method prevented velocity recovery within the first 1-2\,ns (due to the finite time window). By forward-modeling the PDV trace directly, our method removes the need to choose  an STFT window function; effectively extrapolating the velocity behavior, with uncertainty, to the chosen start-time. Skidmore et al.\ also noted that two probes (15 and 23) had unreliable velocity extraction due to spectral broadening resulting from the numerical aperture of the optical fiber. Our approach would be similarly affected by contamination from multiple velocities, but provides a reliably objective method of inferring the signal-to-noise weighted velocity in the fiber.

\begin{figure*}
    \centering
    \includegraphics[width=0.925\textwidth]{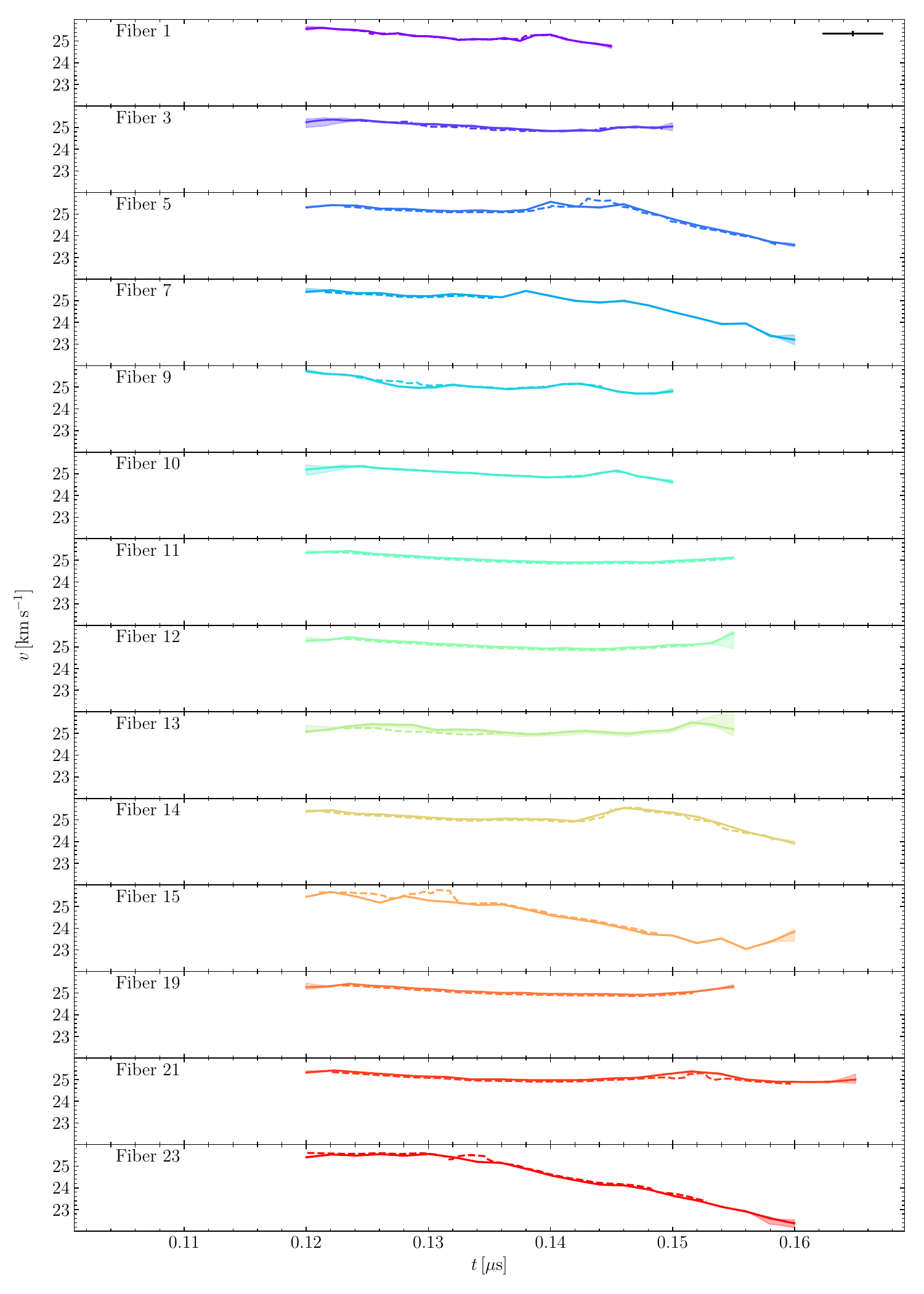}\\
    \caption{Velocity histories (velocity, $v$ versus time, $t$) inferred from experimental PDV data with the STAR 2SLGG. The solid lines and envelopes, respectively, denote the median and 95\,\% credible interval from our Bayesian analysis. The dashed lines denote the analysis by Skidmore et al. using the standard STFT-approach, demonstrating similarity between the two methods. The single black error bar (top right) indicates the velocity uncertainty (0.1275\,km\,s$^{-1}$) and time window (5\,ns) of the STFT-based approach.}
    \label{figure:star_endor_pdv_velocities_pw_model}
\end{figure*}

\subsection{Data preparation}
\label{subsec:data_preparation}

The raw oscilloscope traces were recorded with 50\,GS\,s$^{-1}$ Tektronix digitizers with 25\,GHz bandwidth, which we ingested into a \texttt{Python} dictionary. We removed DC and low-frequency signal components using a high-pass filter with a 2\,GHz cutoff, and then estimated the noise covariance using the technique described in Section\,\ref{subsec:models}. 

To avoid unnecessarily increasing the complexity of our inference problem, we estimated the time-varying behavior of the amplitude empirically from the data. This was achieved by applying a 3$^\mathrm{rd}$-order Savitsky-Golay filter\cite{Savitsky:1964}, with a window size of 10\,ns (several hundred PDV fringes for the expected velocity of 25\,km\,s$^{-1}$), to the absolute values of the PDV data. The data and the noise covariance were then normalised by this estimated amplitude, ideally removing time-varying behavior due to changes in the laser intensities and/or fiber performance. This approach assumes that:
\begin{enumerate}
\item the signal-to-noise ratio is sufficiently high such that the noise does not contribute significantly to our estimate of the amplitude
\item temporal variations in amplitude are significantly slower than the fringe period ($\Delta{t} \gg 0.10$\,ns)
\item the PDV signal contains a single coherent component and not multiple sinusoids
\end{enumerate}
If these assumptions do not hold to desirable tolerances, then the time-varying amplitude will need to be explicitly included in the model, with a significant increase in the computational complexity of the inference. Spectrograms constructed from the resultant normalised oscilloscope traces are shown in the left panels of Fig.\,\ref{figure:star_endor_pdv_spectrograms}.

\renewcommand{\arraystretch}{1.2}
\begin{table}[h]
    \centering
    \caption{Summary of the experimental parameters used for each dataset in our analysis of the STAR 2SLGG shot. Fiber identification numbers (IDs) are as shown in Fig.\ref{figure:star_endor_experimental_setup}. Wavelengths, $\lambda_\mathrm{T}$ and $\lambda_\mathrm{R}$, are for the target and reference lasers, respectively. Fixed start and end times for the velocity history were chosen from visual inspection of the corresponding spectrograms (Fig.\,\ref{figure:star_endor_pdv_spectrograms}).}
    \label{table:experimental_parameters}
    \begin{tabular}{ccccc}
        \hline
        \hline
        Fiber ID & $\lambda_\mathrm{T}$\,[nm] & $\lambda_\mathrm{R}$\,[nm] & Start time\,[$\mu$s] & End time\,[$\mu$s] \\
        \hline
        \hline
        1 & 1549.193 & 1548.923 & 0.120 & 0.145 \\
        3 & 1549.193 & 1548.923 & 0.120 & 0.150 \\
        5 & 1554.125 & 1553.855 & 0.120 & 0.160 \\
        7 & 1554.125 & 1553.855 & 0.120 & 0.160 \\
        9 & 1549.193 & 1548.923 & 0.120 & 0.150 \\
        10 & 1549.193 & 1548.923 & 0.120 & 0.150 \\
        11 & 1550.654 & 1550.383 & 0.120 & 0.155 \\
        12 & 1550.654 & 1550.383 & 0.120 & 0.155 \\
        13 & 1554.125 & 1553.855 & 0.120 & 0.155 \\
        14 & 1554.125 & 1553.855 & 0.120 & 0.160 \\
        15 & 1549.218 & 1548.945 & 0.120 & 0.160 \\
        16 & 1549.218 & 1548.945 & 0.120 & 0.160 \\
        19 & 1550.654 & 1550.383 & 0.120 & 0.155 \\
        21 & 1550.654 & 1550.383 & 0.120 & 0.165 \\
        23 & 1549.218 & 1548.945 & 0.120 & 0.160 \\
        \hline
        \hline
    \end{tabular}
\end{table}

\subsection{Validation results}

We used the method discussed in Section\,\ref{sec:bayesian_pdv_analysis} for analysis of the experimental PDV data, with the parameter priors summarized in Table\,\ref{table:prior_distributions} and the experimental parameters summarized in Table\,\ref{table:experimental_parameters}. In early tests, we found that the sampling algorithm would infer start and end times for the PDV signal that are significantly different to those evident from the spectrogram. This is due to the finite temporal resolution of the parametrized model; a model that is a good fit to a subset of high signal-to-noise data is preferred over one that is a worse fit to the whole dataset. Therefore, in order to recover a model of the whole signal, we fixed the start and end times to those evident from visual inspection of the corresponding spectrograms.

In Fig.\,\ref{figure:star_endor_pdv_velocities_pw_model} we show the inferred velocity histories for each fiber optic probe using our Bayesian analysis and the STFT approach. Qualitatively, the approaches give similar results and our analysis supports the finding of Skidmore et al.\ that the amplifier achieves the required design specifications. Minor systematic velocity differences of $\sim 1$\,\% for some fibers (for example, Fiber\,21) are likely due to details of the extraction. Larger differences in velocity at specific  times (for example, Fiber 15 between 0.125 and 0.135\,$\mu$s) are the result of spectral broadening or laser speckle. The spectrograms generated from the experimental data and those inferred using the model are shown in Fig.\ref{figure:star_endor_pdv_spectrograms}, from which we can see the advantage of the Bayesian approach in being able to infer signal from noisy data (for example, Fiber 3) and interpolate across signal gaps (for example, Fibers 5 and 7). Likewise, the Bayesian approach has allowed us to infer velocity histories at earlier times than was available with the STFT-approach. 

\section{Concluding remarks}
\label{sec:concluding_remarks}

We have presented a methodology for carrying out Bayesian inference of velocity histories directly from PDV oscilloscope data as an adjunct to standard STFT-based analysis. We verified the method using synthetic PDV data generated from a template based on an experiment carried out by Skidmore et al.\cite{Skidmore:2025} with the STAR 2SLGG at Sandia National Laboratories\cite{Chhabildas:1982}. We were able to recover the injected velocity history, within the uncertainties, and the posterior predictive distribution is consistent with the synthetic data. We then applied it to the real experimental data and compared with velocity histories obtained using the standard STFT-based analysis. We find that in general the two approaches agree, although we caution that the prior probability distributions for model parameters need to be carefully chosen to regularize the inference against, for example, aliasing of higher frequency harmonics. The Bayesian approach is found to be particularly advantageous when choices need to be made about interpolating across regions of missing data, low signal-to-noise or laser speckle.

The inferred velocity history is dependent on the model and priors chosen by the user (here we use a piecewise linear model) and, importantly, the resolution. We therefore caution that this does not capture uncertainty due to model misspecification, for example, of any rapidly time-varying or discontinuous behavior. Likewise, multiple velocities can only be captured by either including them explicitly in the model or, possibly, by using multi-modal sampling techniques. Here, we have assumed a single velocity component so that the PDV amplitude can be capture and normalised in a pre-processing step; multiple velocities would require the amplitudes to be explicitly included in the model. These issues can be identified post-hoc using posterior predictive checks (see e.g. Fig.\,\ref{figure:synthetic_verification_posterior_analysis}) and overcome by including a sufficiently flexible model with the required parameters to capture the desired behavior. However, increased model complexity has associated computational costs; even with the model used here the numerical sampling can take several hours to converge. We therefore recommend that it be used as a complementary method to the standard STFT-based approach.

\begin{acknowledgments}
The authors thank the anonymous reviewers for their constructive comments and suggestions, which helped improve the quality of this manuscript. The authors would also like to thank the staff at FLF for supporting the work in this paper, which included provision of the HPC for parallel sampling in our analysis. J.~R. Allison thanks M. Gorman for useful discussions whilst preparing the paper. Sandia National Laboratories is a multi-mission laboratory managed and operated by National Technology and Engineering Solutions of Sandia, LLC (NTESS), a wholly owned subsidiary of Honeywell International Inc., for the U.S. Department of Energy’s National Nuclear Security Administration (DOE/NNSA) under contract DE-NA0003525.  \\
\end{acknowledgments}

\section*{Data Availability Statement}

The data that support the findings of this study are available from the corresponding author upon reasonable request.

\appendix

\section{Demonstration of the noise model}
\label{app:demonstration_of_noise_model}

\begin{figure*}
    \centering
    \includegraphics[width=1.0\textwidth]{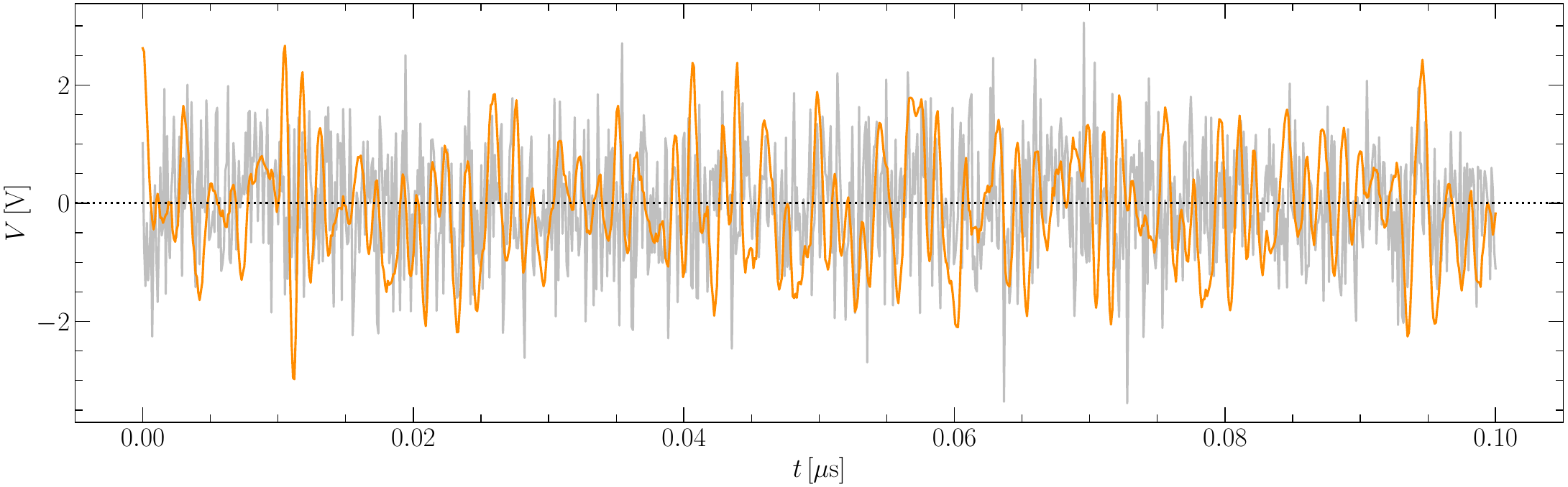}\\
    \caption{Correlated Gaussian noise (orange; voltage, $V$ versus time, $t$) generated using a covariance matrix that models the effect of a 1\,GHz low-pass filter using the method discussed in Section\,\ref{subsubsec:pdv_noise}. Uncorrelated Gaussian noise (grey) is shown for comparison.}
    \label{figure:filtered_noise_pdv_traces}
\end{figure*}

\begin{figure*}
    \centering
    \includegraphics[width=0.95\textwidth]{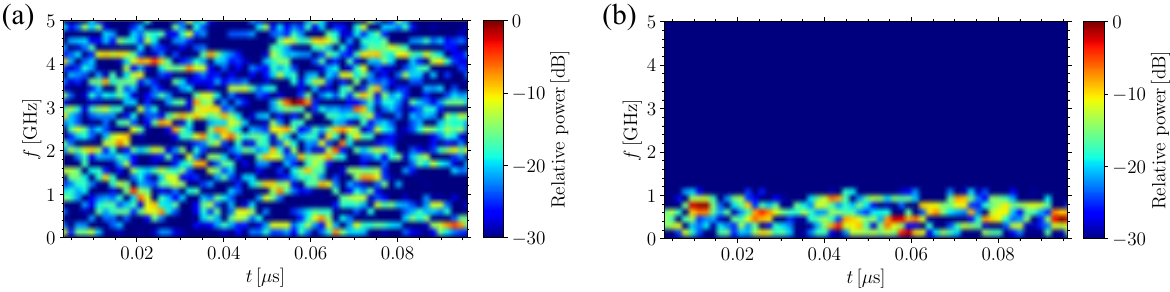}
    \caption{Spectrograms (beat frequency, $f$ versus time, $t$) generated from (a) the uncorrelated and (b) correlated noise-only traces shown in Fig.\,\ref{figure:filtered_noise_pdv_traces}, showing the expected removal of frequencies above 1\,GHz.}
    \label{figure:filtered_noise_pdv_spectrograms}
\end{figure*}

The noise model is a critical component of any inference method. In Section\,\ref{subsubsec:pdv_noise} we described our PDV noise model, which is based on a multivariate Gaussian distribution with mean zero and a covariance matrix estimated from Monte-Carlo simulations of bandpass-filtered white noise. Here we demonstrate the validity of this model; in Fig.\,\ref{figure:filtered_noise_pdv_traces} we show an example of correlated Gaussian noise, sampled at 50\,GHz, that has been generated using this model for a 1\,GHz low-pass filter. The correlations between data at high frequency are visually evident when compared with Gaussian noise that has been generated using a model with no covariance. This is further demonstrated by the spectrograms shown in Fig.\,\ref{figure:filtered_noise_pdv_spectrograms}, which show the successful removal of noise frequencies greater than 1\,GHz.

\section{Experimental validation data}
\label{app:experimental_validation_data}

Experimental validation data were obtained by Skidmore et al. from a shot with the STAR 2SLGG facility at Sandia National Laboratories\cite{Skidmore:2025}. A planar shock front was driven into a quartz sample at about 25\,km\,s$^{-1}$ and PDV data were recorded using fourteen fiber-optic probes in a linear array across a 1\,mm region. The oscilloscope traces were bandpass filtered and the signal amplitude estimated using the method described in Section\,\ref{subsec:data_preparation}. Spectrograms constructed from the resultant normalized oscilloscope traces are shown in the left panels in Fig.\,\ref{figure:star_endor_pdv_spectrograms}. The inferred posterior spectrograms are shown in the right panels of the same figure, highlighting the ability of the Bayesian method to recover the signal from noisy data.

\begin{figure*}[h!]
    \centering
    \includegraphics[width=0.95\textwidth]{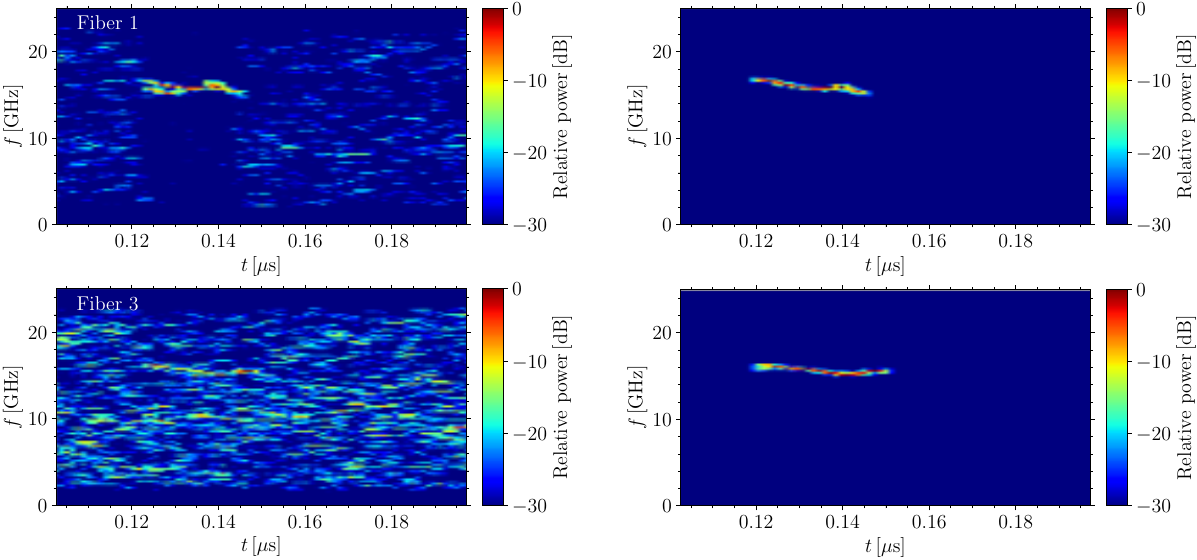}
    \caption{Spectrograms (beat frequency, $f$ versus time, $t$) generated from the experimental validation data discussed in Section\,\ref{sec:experimental_validation} (left panels) and the inferred posterior distribution of model PDV signals (right panels).}
    \label{figure:star_endor_pdv_spectrograms}
\end{figure*}

\addtocounter{figure}{-1}

\begin{figure*}[h!]
    \centering
    \includegraphics[width=0.95\textwidth]{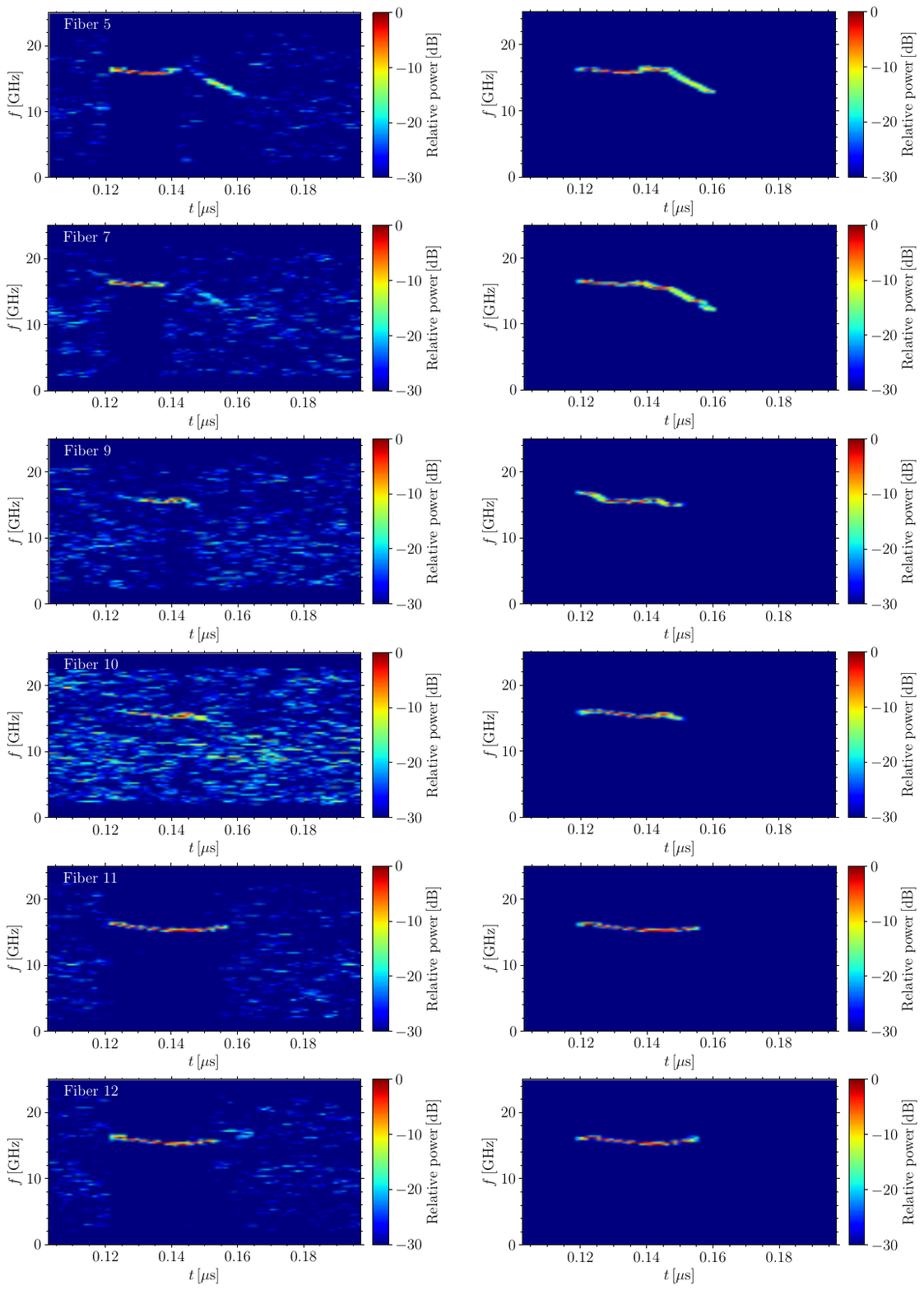}
    \caption{(Continued) Spectrograms (beat frequency, $f$ versus time, $t$) generated from the experimental validation data discussed in Section\,\ref{sec:experimental_validation} (left panels) and the inferred posterior distribution of model PDV signals (right panels).}
\end{figure*}

\addtocounter{figure}{-1}

\begin{figure*}[h!]
    \centering
    \includegraphics[width=0.95\textwidth]{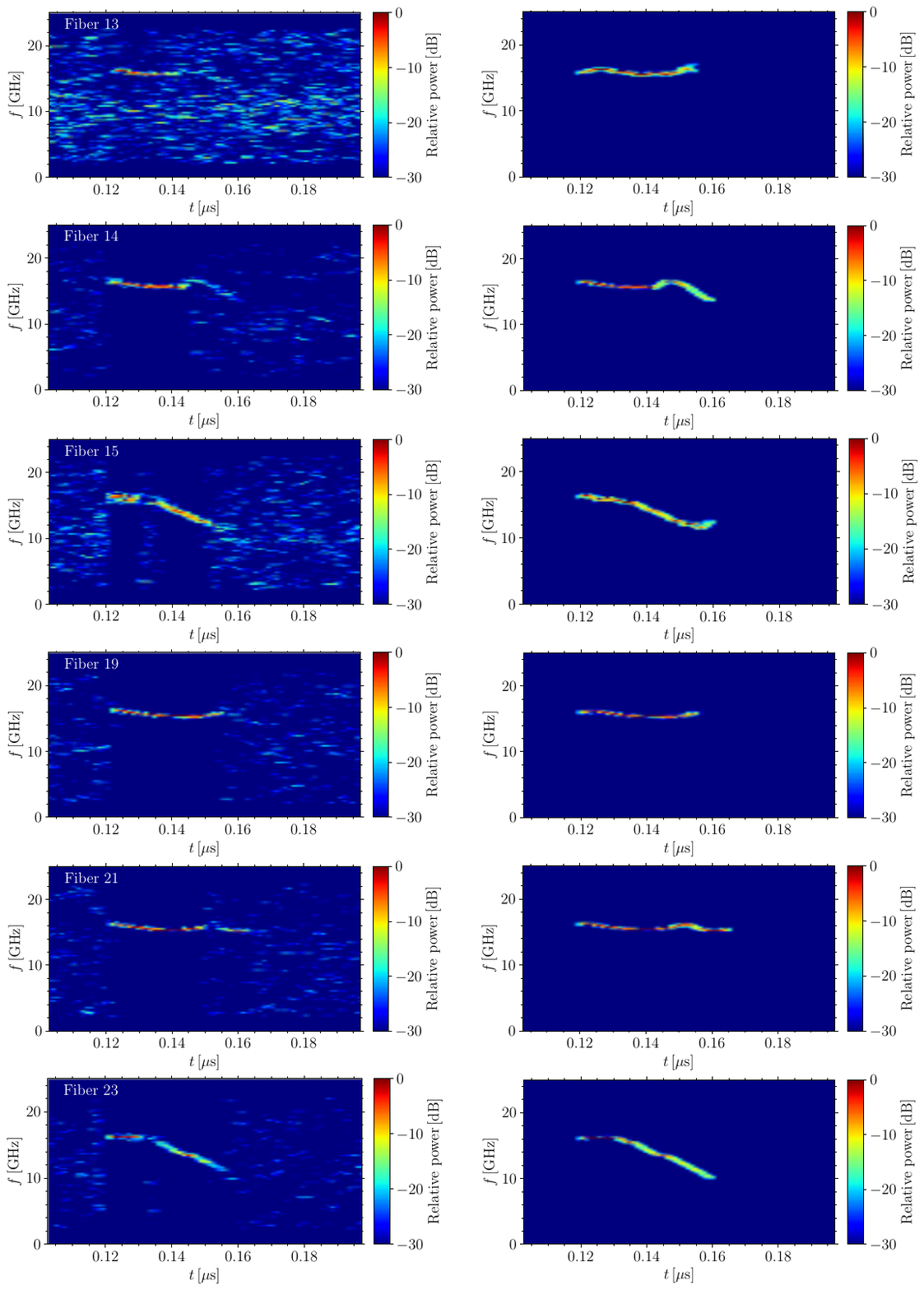}
    \caption{(Continued) Spectrograms (beat frequency, $f$ versus time, $t$) generated from the experimental validation data discussed in Section\,\ref{sec:experimental_validation} (left panels) and the inferred posterior distribution of model PDV signals (right panels).}
\end{figure*}

\nocite{*}
\bibliography{references}

\end{document}